\documentclass[10pt,onecolumn,draftcls]{IEEEtran} 

\usepackage{graphicx,subfigure,slashbox}
\usepackage{amsmath,amsfonts}
\usepackage{color}
\usepackage{url}
\usepackage{cite}

\newtheorem{theorem}{Theorem}[section]

\newcommand{\qed}{~~ \raisebox{.65ex}{\fbox{\rule{0mm}{0mm}}}}

\newcommand{\ie}{{\em i.e.},~}
\newcommand{\eg}{{\em e.g.},~}
\newcommand{\cf}{{\em cf.}~}
\newcommand{\uv}{\underline{v}}
\newcommand{\uone}{\underline{1}}
\newcommand{\uzero}{\underline{0}}
\newcommand{\uy}{\underline{y}}
\newcommand{\utheta}{\underline{\theta}}

\newcommand{\dd}{\mbox{d}}

\newcommand{\beqa}{\begin{eqnarray*}}
\newcommand{\eeqa}{\end{eqnarray*}}
\newcommand{\be}{\begin{eqnarray}}
\newcommand{\ee}{\end{eqnarray}}
\title{Stochastic Loss Aversion for \\
Random Medium Access\thanks{The work 
was supported by NSF CISE grant 0915928
and by a Cisco Systems URP gift.}
}

\author{
G. Kesidi$\mbox{s}^\dagger$ 
and Y. Ji$\mbox{n}^\ddagger$
\\
$\dagger$ CS\&E and EE Depts, Penn State University, gik2@psu.edu
\\
$\ddagger$  EE Dept, KAIST, South Korea, youngmi\_jin@kaist.ac.kr
}

\begin{document}
\maketitle

\begin{abstract}
We consider a slotted-ALOHA LAN with loss-averse,
noncooperative greedy users. To avoid non-Pareto equilibria,
particularly deadlock,
we assume probabilistic loss-averse behavior.
This behavior is modeled as a 
modulated white noise term,
in addition to the greedy term, creating a
diffusion process modeling the game.
We observe that when player's  modulate with
their throughput,  a more efficient exploration of
play-space 
(by Gibbs sampling) 
results, and so finding a Pareto 
equilibrium is more likely over a given interval of
time.
\end{abstract}


\section{Introduction}


The ``by rule" 
window flow control mechanisms of,
\eg TCP and CSMA, have elements of
both proactive and reactive communal
congestion control suitable
for distributed/information-limited high-speed networking scenarios.
Over the past ten years, 
game theoretic models for medium access and flow control 
have been extensively explored in order to consider the
effects of even a single  end-user/player who greedily departs from such
prescribed/standard behaviors
\cite{Jin02a,Jin02b,Wicker03,Jin05,Cagalj05,Altman06,Menache07,Lee07,Cui08,Ma09,JinKes12}.
Greedy end-users may have a dramatic effect on the
overall ``fairness" of the communication network under consideration.
So, if even one end-user acts in a greedy way, it may be
prudent for all of them to do so.
However, even end-users with an noncooperative disposition may
temporarily not practice greedy behavior
in order to escape from sub-optimal (non-Pareto) Nash equilibria.
In more general game theoretic contexts, the reluctance of an
end-user to act in a non-greedy fashion is called
loss aversion  \cite{loss-aversion}.

In this note, we focus on simple slotted-ALOHA MAC for a LAN. 
We begin with a noncooperative model of end-user behavior.
Despite the presence of a stable interior Nash equilibrium,
this system was shown in \cite{Jin02a,Jin02b} to have a large
domain of attraction to deadlock where all players' transmission
probability is one  and so obviously all players' throughput is zero
(here assuming feasible  demands and throughput based costs).
To avoid non-Pareto Nash equilibria, particularly those  involving
zero throughput for some or all users, we assume that end-users will
{\em probabilistically} engage in non-greedy behavior. That is,
a stochastic model of loss aversion, a behavior  whose aim is long term 
communal betterment.

We may be able to model a play that reduces net-utility using
a single ``temperature" parameter $T$ 
in the manner of simulated annealing
(\eg \cite{Holley-Stroock}); \ie plays that increase net utility are always
accepted and plays that reduce net utility are (sometimes) accepted with
probability decreasing in $T$, so the players are
(collectively) less loss averse with larger $T$.
Though our model of
probabilistic loss aversion is related that of simulated
annealing by diffusions \cite{Gidas85,Wong91}, even with a
free meta-parameter 
($\eta$ or $\eta w$ below) 
possibly interpretable as temperature,
our modeling aim is not centralized annealing 
(temperature cooling) rather decentralized exploration
of play-space by noncooperative users.

We herein do not model how the end-users will keep
track of the best (Pareto) equilibria previously 
played/discovered\footnote{The players could, \eg
alternate between (loss averse) greedy behavior to discover Nash
equilibrium points, and the play dynamics modeled herein
for breadth of search (to escape non-Pareto equilibria).}.
Because 
the global extrema of the global objective functions
(Gibbs exponents) we derive do not 
necessarily correspond to Pareto equilibria,
we do not advocate collective slow
``cooling" (annealing) of the equivalent temperature parameters.
Also, we do not model how end-user throughput demands may be time-varying,
a scenario which would motivate the ``continual search" aspect
of the following framework. 

The following
stochastic approach to distributed play-space search is  also
related to ``aspiration" of repeated games \cite{Bendor01,Cho05,KMRV98},
where
a play resulting in suboptimal utility may be accepted
when the utility is less than a threshold, say according
to a ``mutation" probability \cite{Kandori93,Montanari09}.
This type of
``bounded rational" behavior been proposed 
to find Pareto equilibria, in particular for 
 distributed settings where players act with limited
information \cite{Montanari09}.
Clearly, given a global
objective $L$ whose global maxima correspond to Pareto equilibria,
these ideas are similar to the use of simulated annealing
to find the global maxima of $L$ while avoiding suboptimal local
maxima.

This paper is organized as follows. In Section \ref{formulation-sec},
we formulate the basic ALOHA noncooperative game under consideration.
Our stochastic framework (a diffusion) for loss aversion is 
given in Section \ref{la-sec}; for
two different modulating terms of the white-noise process,
the invariant distribution in the collective play-space is derived.
A two-player numerical example is used to illustrate the
performance of these
two approaches in Section \ref{numerical-sec}.
We conclude in Section \ref{concl-sec} with a discussion
of future work.

\section{A distributed slotted-ALOHA game for LAN MAC}\label{formulation-sec}

Consider an idealized\footnote{We herein do not consider physical layer
channel phenomena such as shadowing and fading as in,
\eg \cite{Menache07,JinKes12}.}
 ALOHA LAN where
each user/player  $i\in\{1,2,...,n\}$  has (potentially different) 
transmission
probability $v_i$.  For the collective 
``play" $\uv=(v_1,v_2,...,v_n)$, 
the net utility of player $i$ is
\be\label{net-util-def}
V_i(\uv) & = & U_i(\theta_i(\uv)) - M\theta_i(\uv),
\ee
where the strictly convex and increasing utility $U_i$ of 
steady-state throughput
\beqa
\theta_i & := & v_i\prod_{j\not=i}(1-v_j) 
\eeqa
is such that $U_i(0)=0$, and
the throughput-based price is $M$.
So, the throughput-demand of the $i^{{\rm th}}$ player is
\beqa
y_i  & := & (U')^{-1}(M).
\eeqa
This is a quasi-stationary game wherein future action is based
on the outcome of the current collective play $\uv$ observed
in steady-state \cite{Brown51}. 

The corresponding continuous
Jacobi iteration of the better response dynamics 
is \cite{Jin02a,Jin02b,Shamma05}: 
for all $i$
\be\label{jacobi-iter}
\frac{\dd}{\dd t} v_i & = & \frac{y_i}{\prod_{j\not=i} (1-v_j)} -v_i
~ =: ~ -E_i(\uv),
\ee
\cf (\ref{dt-game}).
Note that we define $-E_i$, instead of $E_i$, to be consistent
with the notation of \cite{Wong91}, which seeks to minimize a global
objective, though we want to  maximize such objectives in the following.

Such dynamics generally exhibit multiple 
Nash equilibria, including  non-Pareto equilibria with significant
domains of attraction. Our ALOHA context has a 
stable deadlock equilibrium point where all players always transmit,
\ie $\uv=\uone:=(1,1,...,1)$ \cite{Jin02a,Jin02b}.

\section{A diffusion model of loss aversion}\label{la-sec}

Generally in the following, we consider {\em differently} loss-averse players.
Both examples considered
are arguably {\em distributed} (information limited) games 
wherein
every player's choice of transmission probability is 
based on information knowable
to them only through their channel observations, 
so that consultation among users is not required.
In particular, players are not directly aware of each other's demands ($y$).

\subsection{Model overview}

We now model 
stochastic perturbation of the Jacobi dynamics
(\ref{jacobi-iter}), allowing for suboptimal
plays despite loss aversion, together with a sigmoid
mapping $g$ to ensure plays (transmission
probabilities) $\uv$ remain in a feasible
hyperrectangle  $D\subset [0,1]^n$ (\ie the feasible
play-space for $\uv$): for all $i$,
\be
\dd u_i & = & -E_i(\uv)\dd t + \sigma_i(v_i)\dd W_i \label{la-u}\\
v_i & = & g_i(u_i)\label{la-v}
\ee
where
$W_i$ are independent standard Brownian motions.
An example sigmoid  is
\be\label{sigmoid-def}
g(u) & := & \gamma(\tanh(u/w)+\delta),
\ee
where $1\leq \delta < 2$ and $0<\gamma \leq 1/(1+\delta)$.  Thus, 
$\inf_u g(u)=\inf v=\gamma(-1+\delta)\geq 0$ and 
$\sup_u g(u)=\sup v =\gamma(1+\delta)\leq 1$.
Again, to escape from the domains of attraction of non-Pareto equilibria,
the deterministic Jacobi dynamics (\ie $-E_i(\uv)\dd t$ in (\ref{la-u}))
have been perturbed by white  noise ($\dd W_i$) here modulated by 
a diffusion term of the form:
\beqa
\sigma_i(v_i) & = &\sqrt{\frac{2h_i(\uv)}{f_i(v_i)}},
\eeqa
where 
\beqa
f_i(v_i)  & := &  g_i'(g_i^{-1}(v_i)).
\eeqa
For  the example sigmoid (\ref{sigmoid-def}),
\beqa
f(v) & = &\frac{\gamma}{w}\left(1-\left(\frac{v}{\gamma}-\delta\right)^2\right).
\eeqa
In the following, we will consider different functions $h_i$
leading to Gibbs invariant distributions for $\uv$.

Note that the discrete-time ($k$) version of this game model would be 
\be
u_i(k+1)-u(k) & = & -E_i(\uv(k))\varepsilon + \sigma_i(\uv(k)) N_i(k)
\nonumber \\
v_i(k+1) & = & g_i(u_i(k+1)), 
\label{dt-game} 
\ee
where the $N_i(k)$ are all i.i.d. normal ${\sf N}(0,\varepsilon)$ random variables.

The system just described is 
a variation of E. Wong's diffusion machine 
\cite{Wong91},  
the difference being the introduction of the
term $h$ 
instead of a temperature meta-parameter $T$.
Also, the diffusion function $\sigma_i$ is player-$i$ dependent
at least through $h_i$. Finally, under the slotted-ALOHA dynamics,
there is no function $E(\uv)$ 
such that $\partial E/\partial v_i = E_i$, so 
we will select the diffusion factors $h_i$ 
to achieve a tractable Gibbs stationary distribution
of $\uv$, and interpret them in terms of  player
loss aversion.

Note that in the diffusion machine, a common temperature parameter $T$ may 
be slowly reduced to zero to find the minimum of
a global potential function  (the exponent of the Gibbs 
stationary distribution of $\uv$)
\cite{TNN95,Kesidis09}, in the manner of simulated
annealing. Again, the effective temperature parameter here
($\eta$ or $\eta w$) will be constant.


\subsection{Example diffusion term $h_i$ decreasing in $v_i$}

In this subsection, we analyze the model when, for all $i$,
\be\label{decr-h}
h_i(v_i)  & :=& \eta y_i(1-v_i)^2.
\ee
with $\eta>0$ a free meta-parameter (assumed common to all players).
So, a greedier player $i$ (larger $y_i$)
will generally tend to be less loss averse (larger $h_i$), except when
their current retransmission play $v_i$ is large.

~\\
\begin{theorem}\label{decr-h-thm}
The  stationary probability density function  of 
$\uv\in D\subset [0,1]^n$, defined by
(\ref{la-v}) and (\ref{la-u}), 
is
\be\label{gibbs}
p(\uv) &  = &  
\frac{1}{Z}
\exp\left(\frac{\Lambda(\uv)}{\eta Y}-\log H(\uv)\right),
\ee
where:  the normalizing term
\beqa
Z  & := &   \int_{D}
\exp\left(\frac{\Lambda(\uv)}{\eta Y}-\log H(\uv)\right)
\dd \uv,\\
D & := & \prod_{i=1}^n \left(\gamma_i(-1+\delta_i),~\gamma_i(1+\delta_i)\right)\\
\Lambda(\uv) & := & \prod_{i=1}^n 
\frac{y_i}{1-v_i}
-  \sum_{j=1}^N \left(
\frac{v_j}{1-v_j} + \log(1-v_j)
\right)
\prod_{i\not=j} y_i\\
H(\uv)  & :=  & \prod_{j=1}^n(1-v_i)^2, ~\mbox{and} \\
Y  & := &  \prod_{j=1}^n y_j.
\eeqa
\end{theorem}

~\\ 
{\em Remark:}
$\Lambda$ is 
a Lyapunov function of 
the deterministic ($\sigma_i\equiv 0$ for all $i$) Jacobi iteration  
\cite{Jin02a,Jin02b}.

~\\
\IEEEproof
Applying Ito's lemma \cite{Karatzas-Shreve,Wong91} to 
(\ref{la-u}) and (\ref{la-v}) gives
\beqa
\dd v_i & = &  g_i'(u_i) \dd u_i + \frac{1}{2} g_i''(u_i)\sigma_i^2(\uv)\dd t\\
& = & 
[-f_i(v_i)E_i(\uv) + \frac{1}{2} g_i''(g_i^{-1}(v_i))\sigma_i^2(\uv)]\dd t\\
 & & ~~~+ f_i(v_i)\sigma_i(\uv)\dd W_i,
\eeqa
where the derivative operator $z'  :=  \tfrac{\dd}{\dd v_i} z(v_i)$
and we have just substituted (\ref{la-u}) for the second equality.
From the Fokker-Planck (Kolmogorov forward) equation for this diffusion
\cite{Karatzas-Shreve,Wong91},
we get the following equation for the time-invariant (stationary)
distribution $p$
of $\uv$: for all $i$,
\beqa
0& = & \frac{1}{2}
\partial_i
(f_i^2\sigma_i^2 p)
-[-f_iE_i + \frac{1}{2} 
(g_i''\circ g_i^{-1})\sigma_i^2]p,
\eeqa
where the operator $\partial_i :=  \tfrac{\partial}{\partial v_i} $.

Now note that 
\beqa
f_i^2(v_i)\sigma_i^2(\uv) & = & 2h_i(v_i)f_i(v_i) ~~\mbox{and}\\
g_i''(g_i^{-1}(v_i))\sigma_i^2(v_i) & = & h_i(v_i) g_i''(g_i^{-1}(v_i))/f_i(v_i) \\
& = & h_i(v_i)f_i'(v_i).
\eeqa
So, the previous display reduces to
\beqa
0 & = & 
\partial_i 
(h_i f_i p) - (-E_i f_i + h_i f_i')p \\
& = & 
(h_i  \partial_i p+ h_i'  p + E_i  p)f_i,
\eeqa
where the second equality is due to cancellation
of the $h_if_i'p$ terms.
For all $i$, since $f_i > 0$,
\be
\frac{\partial_i p(\uv)}{p(\uv)}=\partial_i \log p(\uv) & = & 
- \frac{E_i(\uv)}{h_i(v_i)} - \frac{h_i'(v_i)}{h_i(v_i)}
\label{logp-equ}\\
& = & \frac{1}{\eta Y}\partial_i \Lambda(\uv)   +\frac{2}{1-v_i}.
\nonumber
\ee
Finally, (\ref{gibbs}) follows by direct integration.
\qed

~\\
Unfortunately,
the exponent of $p$ under (\ref{decr-h}), 
\beqa
\tilde{\Lambda}(\uv)  & :=&   \frac{\Lambda(\uv)}{\eta Y}-\log H(\uv),
\eeqa
and both its component terms 
$\Lambda$ and $-\log H$,
remain maximal in the deadlock region near $\uone$.

\subsection{Example diffusion term $h_i$ increasing in $v_i$}

The following alternative diffusion term $h_i$ 
is an example which is instead
increasing in $v_i$, but decreasing  in the 
channel {\em idle time} from player $i$'s point-of-view
\cite{Heusse05,Antoniadis11}, 
\be\label{idle-h}
h_i(\uv) & := & \frac{\eta v_i}{\prod_{j\not=i}(1-v_j)}.
\ee
That a user would be less loss averse (higher $h$) when
the channel was perceived to be more idle may be
a reflection of a ``dynamic" altruism \cite{Antoniadis11}
(\ie  a player is more courteous as s/he perceives that others
are).
The  particular form of 
(\ref{idle-h}) also leads to another
tractable Gibbs distribution for $\uv$.

~\\
\begin{theorem}
Using (\ref{idle-h}), the stationary probability density function
of  the diffusion $\uv$ on $[0,2\gamma]^n$
is
\be\label{gibbs2}
p(\uv) &  = &  \frac{1}{W} \exp(\Delta(\uv))
\ee
where 
\be\label{Delta}
\Delta(\uv) & = & 
\sum_{i=1}^n\left(\frac{y_i}{\eta}-1\right)\log v_i
+ \frac{1}{\eta}\prod_{i=1}^n(1-v_i),
\ee
and $W$ is the normalizing term.
\end{theorem}

~\\
\IEEEproof
Following the proof of Theorem \ref{decr-h-thm},
the invariant here satisfies also satisfies (\ref{logp-equ}):
\beqa
\partial_i \log p(\uv) & = & 
- \frac{E_i(\uv)}{h_i(\uv)} - \partial_i \log h_i(\uv)\\
& = & \frac{y_i}{\eta v_i} - \frac{1}{\eta}\prod_{j\not=i}(1-v_j) -
\frac{1}{v_i}.
\eeqa
Substituting (\ref{idle-h}) gives:
\beqa
\partial_i \log p(\uv) & = & 
\left(\frac{y_i}{\eta}-1\right)
\frac{1}{v_i} - \frac{1}{\eta}\prod_{j\not=i}(1-v_j).
\eeqa
So, we obtain (\ref{Delta}) by direct integration.
\qed

~\\
Note that if $\eta>\max_i y_i$, then $\Delta$ is 
strictly decreasing in $v_i$ for all $i$, and
so will be minimal in the deadlock region (unlike
$\tilde{\Lambda}$).  So the
stationary probability in the region of deadlock 
will be low. However, large $\eta$ may result
in the stationary probability close to $\uzero$
being very high. 
So, we see that the meta-parameter
$\eta$ (or $\eta w$) here plays a more significant role (though
the parameters $\delta$ and $\gamma$ in $g$ play a more significant
role in the former objective $\tilde{\Lambda}$ owing to its 
global extrema at $\uone$).

\section{Numerical examples}\label{numerical-sec}

\subsection{Using (\ref{decr-h})}


For an $n=2$ player example with demands
$\uy=(8/15,1/15)$ and $\eta=1$, the 
two interior Nash equilibria are
the locally stable (under deterministic dynamics) 
at $\uv^*_a=(2/3,1/5)$ 
and the (unstable) saddle point at  $\uv^*_b=(4/5,1/3)$  
(both with corresponding throughputs $\utheta = \uy$) \cite{Jin02a,Jin02b}. 
Again, $\uone$ is a stable deadlock boundary
equilibrium which is naturally to be avoided if possible
as both players' throughputs are zero there, $\utheta=\uzero$.
Under the deterministic dynamics of (\ref{jacobi-iter}), the
deadlock equilibrium $\uone$ had a significant domain of attraction
including a neighborhood of the saddle point $\uv^*_b$.

\begin{figure}[ht]
\includegraphics[width=3.25in]{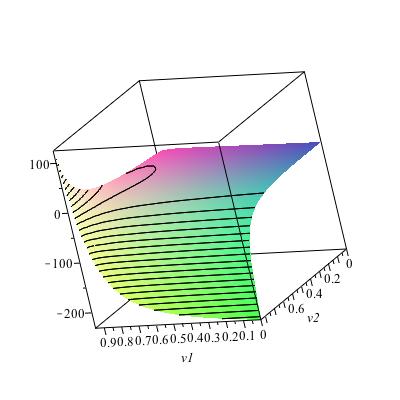} 
\caption{The Gibbs distribution (\ref{gibbs}) for $n=2$ players
with demands $\uy = (8/15,1/15)$ under (\ref{decr-h})}\label{decr-h-fig}
\end{figure}

The exponent of $p$, $\tilde{\Lambda}$, for this example is
depicted in Figure \ref{decr-h-fig}. $\tilde{\Lambda}$ has  a similar
shape as that the Lyapunov function $\Lambda$, 
but without interior local extrema
or saddle points. The extreme mode at $\uone$ is clearly evident.

When we took $\gamma_i,\delta_i$ in (\ref{sigmoid-def}) so
that $0.05 \leq v_i=g(u_i)\leq 0.85$ for all players $i$, 
the stationary probability of the
``good" region containing the
two interior Nash equilibria is 
\be\label{decr-perf}
{\sf P}(\uv\in[0.65,0.82]\times[0.18,0.35]) &\approx & 0.18,
\ee
as computed using (\ref{gibbs}). 
This probability does not appreciably improve by varying
$\eta$ from 1 (both $\Lambda$ and $\log H$ diverge at $\uone$), 
though it does dramatically decrease as
$\sup_u g(u) \uparrow 1$, \eg if
we take 
$\sup_u g(u)= 0.9$  then
(\ref{decr-perf}) decreases to about $10^{-54}$
(essentially zero, of course) which is consistent with Figure \ref{decr-h-fig}.
Also, the largest contribution of this probability is the region around
the saddle point which is part of the deadlock domain of attraction of
$\uone$, the global maximum of $\tilde{\Lambda}$.

To reiterate,
the fundamental advantage of stochastic   
loss aversion is seen by 
 comparing the large domain of attraction
of the deadlock equilibrium 
of the deterministic 
dynamics \cite{Jin02b}, with positive probability 
of presence near the interior Nash equilibrium points
where both players' demands $y$ are satisfied.
This advantage is born out more clearly in the following
example.

%
%

\subsection{Using (\ref{idle-h})}

For the same two-player example
with $\eta=4.5/15=(y_1+y_2)/2$,
and $\sup_u g(u)=0.9$, the probability in (\ref{decr-perf})
was 0.05,
compared to essentially zero for the same parameter range for
(\ref{decr-h}). 
The exponent of $p$, $\Delta$, for this example is
depicted in Figure \ref{decr-h-fig}; $\Delta$  is dissimilar to
the function $\tilde{\Lambda}$ (or $\Lambda$ or $H$) 
without significant modes in the
range of $\eta$ close to the demands $y$, and a much smaller
overall range of values (and likewise for the stationary
density $p$ of $\uv$). 
Though the performance under (\ref{idle-h}) was more sensitive to 
$\eta$ (temperature) than under (\ref{decr-h}),  using (\ref{idle-h}) 
clearly resulted in more effective searching of the play-space 
$D$ and was
far less sensitive to the parameters $\delta,\gamma$ defining it.

\begin{figure}[ht]
\includegraphics[width=3.25in]{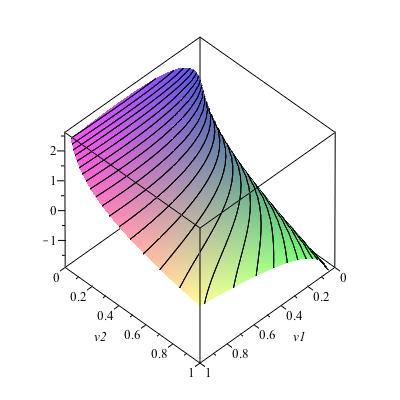} 
\caption{The Gibbs distribution (\ref{gibbs2}) for $n=2$ players
with demands $\uy = (8/15,1/15)$ under (\ref{idle-h})}\label{idle-h-fig}
\end{figure}


\section{Conclusions and Future Work}\label{concl-sec}

The diffusion term (\ref{idle-h}) was clearly
more effective than (\ref{decr-h}) at exploring the play-space, and 
in so doing, was dramatically less sensitive to the choice of
the parameters $\delta$ and $\gamma$ governing the 
range of the play-space $D$.

In future work, we plan to explore other diffusion factors $h$ 
(numerically if they do not lead to a Gibbs stationary distribution $p$)
with a goal to
reduce the stationary probability that $\uv$ occupies the boundary regions.
Also, we will consider a model with power based costs, \ie
$Mv$ instead of $M\theta$ in the net utility (\ref{net-util-def}). Finally,
we will study the effects of asynchronous and/or multirate play among the users
\cite{Bertsekas89,Jin05,Antoniadis11}.


\begin{thebibliography}{9}

\bibitem{Altman06}
 E. Altman, T. Boulogne,  R. El-Azouzi, T. Jim\'{e}nez and L. Wynter.
 A survey on networking games in telecommunications.
 {\em Comput. Oper. Res.}
 {\bf 33}(2):286--311, 2006.

\bibitem{Antoniadis11}
P. Antoniadis, S. Fdida, C. Griffin, Y. Jin, G. Kesidis.
CSMA Local Area Networking under Dynamic Altruism.
{\em submitted}, Dec. 2011.

\bibitem{Bendor01}
J. Bendor, D. Mookherjee and B. Ray. Aspiration-based reinforcement
learning in repeated interaction games: an overview. {\em International Game
Theory Review} {\bf 3}(2\&3): pp. 159–174, 2001.

\bibitem{Bertsekas89}
D.P. Bertsekas and J.N. Tsitsiklis.
Convergence rate and termination of asynchronous iterative algorithms.
In {\em Proc. 3rd International Conference on Supercomputing}, 1989.

\bibitem{Brown51}
G.W. Brown. 
\newblock Iterative solutions of games with fictitious play.
\newblock In {\em Activity Analysis of Production 
and Allocation}, T.C. Koopmans (Ed.), Wiley, New York, 1951.

\bibitem{Cagalj05}
M. Cagalj, S. Ganeriwal, I. Aad and J.P. Hubaux.
On Selfish Behavior in CSMA/CA networks.
In {\em Proc. IEEE INFOCOM}, 2005.

\bibitem{loss-aversion}
C.F. Camerer and G. Loewenstein.  Behavioral Economics: Past, Present, Future.
In {\em Advances in Behavioral Economics},
C.F. Camerer, G. Loewenstein and M. Rabin (Eds.),
Princeton Univ. Press, 2003.

\bibitem{Cho05}
I.-K. Cho and A. Matsui. Learning aspiration in repeated games.
{\em Journal of Economic Theory} {\bf 124}: pp. 171–201, 2005.

\bibitem{Cui08}
T. Cui, L. Chen, and S.H. Low.
A Game-Theoretic Framework for Medium Access Control.
{\em IEEE Journal on Selected Areas in Communications}
{\bf 26}(7), Sept. 2008.

\bibitem{Gidas85}
B. Gidas. Global optimization via the Langevin equation.
In {\em Proc. IEEE CDC}, Ft. Lauderdale, FL, Dec. 1985.

\bibitem{Heusse05}
M. Heusse, F. Rousseau, R. Guillier and A. Dula.
Idle sense: An optimal access method for high throughput and fairness in
rate diverse wireless LANs.
In {\em Proc. ACM SIGCOMM}, 2005.

\bibitem{Holley-Stroock}
R. Holley and D. Stroock.
Simulated Annealing via {S}obolev Inequalities.
{\em Communications in Mathematical Physics}
{\bf 115}(4), Sept. 1988.

\bibitem{Jin02a}
Y. Jin and G. Kesidis.
A pricing strategy for an {ALOHA} network of heterogeneous users with inelastic bandwidth requirements.
In {\em Proc. CISS, Princeton}, March 2002.

\bibitem{Jin02b}
Y. Jin and G. Kesidis.
Equilibria of a noncooperative game for heterogeneous users of an {ALOHA} network.
{\em IEEE Communications Letters} {\bf  6}(7):282-284, 2002.
	
\bibitem{Jin05}
Y. Jin and G. Kesidis.
Dynamics of usage-priced communication networks:
the case of a single bottleneck resource.
{\em IEEE/ACM Trans. Networking}, Oct. 2005.


\bibitem{JinKes12}
Y. Jin and G. Kesidis.
A channel-aware MAC protocol in an ALOHA
network with selfish users.
{\em IEEE JSAC Special Issue on Game Theory in Wireless Communications},
Jan. 2012.

\bibitem{Kandori93}
M. Kandori, G. Mailath, and R. Rob. Learning, mutation, and long run
equilibria in games. {\em Econometrica} {\bf 61}(1): 29–56, Jan. 1993.

\bibitem{KMRV98}
R. Karnadikar, D. Mookherjee, D. Ray, and F. Vega-Redondo. Evolving
aspirations and cooperation. {\em Journal of Economic Theory} {\bf 80}: pp.
292–331, 1998.

\bibitem{Karatzas-Shreve}
I. Karatzas and S.E. Shreve.
{\em Brownian Motion and Stochastic Calculus}.
Springer, 1991.

%

\bibitem{TNN95}
G. Kesidis.  Analog Optimization with Wong's Stochastic
Hopfield Network.  {\em IEEE Trans.  Neural Networks}
{\bf 6}(1), Jan. 1995.

\bibitem{Kesidis09}
G. Kesidis.
A quantum diffusion network.
Technical Report 0908.1597, 2009, 
available at http://arxiv.org/abs/0908.1597


\bibitem{Lee07}
J.W. Lee, M. Chiang, and R.A. Calderbank.
Utility-optimal random-access protocol.
{\em IEEE Transactions on Wireless Communications}
{\bf 6}(7), July 2007.

\bibitem{Ma09}
R.T.B. Ma, V. Misra, and D. Rubenstein.
An Analysis of Generalized Slotted-Aloha Protocols.
{\em IEEE/ACM Transactions on Networking}
{\bf 17}(3), June 2009.

\bibitem{Menache07}
I. Menache and N. Shimkin. 
Fixed-rate equilibrium in wireless collision channels.
In {\em Proc. Network Control and Optimization (NET-COOP)}, Avignon,
France, June 2007.

\bibitem{Montanari09}
A. Montanari and A. Saberi. Convergence to equilibrium in local
interaction games. In {\em FOCS}, 2009. 


\bibitem{Shamma05}
J.S. Shamma and G. Arslan. 
Dynamic fictitious play, dynamic gradient play, and distributed 
convergence to Nash equilibria. 
{\em IEEE Trans. Auto. Contr.} {\bf 50}(3):312-327, 2005.

\bibitem{Wicker03}
S.B. Wicker  and A.B. MacKenzie.
Stability of Multipacket Slotted Aloha with Selfish Users and Perfect 
Information.
In {\em Proc. IEEE INFOCOM}, 2003.

\bibitem{Wong91}
E. Wong. Stochastic Neural Networks. {\em Algorithmica} {\bf 6}, 1991.

\end{thebibliography}
\bibliographystyle{plain}

\end{document}